\documentclass[aps,prb,groupedaddress,amsmath,amssymb,showpacs]{revtex4}

\usepackage{graphicx}
\usepackage{dcolumn}
\usepackage{bm}

\begin{document}


\title{Electron localization in linear chains of identical loop scatterers}


\author{Chih-Lung Chou}
\email{choucl@cycu.edu.tw}

\affiliation{Department of Physics, Chung Yuan Christian University,\\
Taoyuan 32023, Taiwan} \affiliation{Department of Science
Education, Taipei Municipal
University of Education,\\
Taipei 10048, Taiwan}


\date{\today}

\begin{abstract}
We show that electron localization is generic in a linear chain of
identical simple quantum wire loops with equal arm lengths in the
presence of either a perpendicular magnetic field or the
spin-orbit interaction, and has less to do with the shapes of the
loops. We calculate the transfer matrices for a general simple
loop scatterer in the presence of these effects. Based on the
knowledge of the transfer matrices, we thus provide a criterion
for the occurrence of the localization and present a simple
formalism to integrate the transmission probability over the
injection wave vector of electron.
\end{abstract}


\maketitle


\section{Introduction\label{sec:intro}}
It has been shown that electron localization can be induced by
magnetic field in particular class of two dimensional quantum
network called the $\mathcal{T}_3$ network\cite{ABcage}. The
localization is due to the interplay of the network geometry and
the Aharonov-Bohm effect\cite{ABeffect}. In this case, electrons
travelling in the network will acquire quantum phases due to the
Aharonov-Bohm effect. For specific values of the magnetic field,
totally destructive interference occurs and the electron motion is
confined in a small portion of the network that is called AB cage.
In experiment the onset of electron localization in a quantum
network is usually detected by transport measurement. This
localization phenomenon has been experimentally observed in
superconduction $\mathcal{T}_3$ network\cite{ABcageSuper} and in
two-dimensional normal metal networks\cite{ABcageMetal}. Different
aspects of AB cages such as the effect of disorder\cite{disorder},
interaction induced delocalization\cite{delocalization}, and
transport\cite{transport} are also discussed in literature.
Electron localization is also predicted in a linear chain of
square loops connected at one vertex in the presence of the
spin-orbit interaction called the Rashba SO
coupling\cite{RashbaNetwork,RashbaNetDetail}. Similarly, the
localization is due to the interplay of the network geometry and
the Aharonov-Casher effect\cite{ACeffect}. In this case, electrons
travelling in the network will acquire quantum phases due to the
Aharonov-Casher effect. For specific values of the spin-orbit
coupling strength, electrons are forbidden to transport in the
chain and the averaged conductance becomes zero.

In this paper we discuss when and why the electron localization
phenomenon occurs in a linear chain of $N$ identical scatterers in
the large $N$ limit. If the scatterers are quantum wire loops,
will the shape of the loops matter in the occurrence of the
localization phenomenon? To answer these questions, we first
review the parametrization of the transfer matrix for a single
scatterer with a single input lead and a single output lead
connected to the scatterer. The electron transmission probability
for a linear chain of the scatterers can be easily obtained from
the elements of the transfer matrix. Typically, the transmission
probability reveals a binary behavior. In some domains of the
parameters that parameterize transfer matrix, the transmission
probability vanishes in the large $N$ limit. In the other domains
of the parameters, the transmission probability is relatively high
and oscillates rapidly. In real situation, the binary structure of
electron transmission may be determined by the factors such as the
injection electron wave vector $k$, the geometry of the scatterer,
and other physical effects such as magnetic field and spin-orbit
interaction, et.c. It is possible that zero electron transmission
always occurs in some linear chains of scatterers no matter what
the value of $k$ is. In this case, electron localization is seen
as zero transmission in the averaged transmission probability
after integrating over the injection wave vector $k$.

The paper is organized as follows. In section \ref{sec:scatter},
we review the parametrization of the transfer matrix for a
spin-blind scatterer with an input lead and an output lead
connected to it. The transfer matrix is known to respect $U(1,1)$
symmetry. We then discuss the transmission probability for the
linear chain of the scatterers and express the probability in
terms of the parameters that parameterize the transfer matrix. We
also define a criterion function for determining whether the
electron transmission is forbidden in the chain of scatterers. We
show that the criterion function can be easily read off from the
transfer matrix of a single scatterer. On the other hand, a
typical spin-sensitive scatterer may respect a larger symmetry,
the $U(2,2)$ symmetry. In some cases, however, the transfer matrix
of a spin-sensitive scatterer is factorized into the spatial part
and the spin part, and respects $U(1,1)\otimes U(2)$ symmetry.
Here $U(1,1)$ is the symmetry for the spatial part of the transfer
matrix and $U(2)$ is the symmetry for the spin part of the
transfer matrix. In these cases, the total transmission
probability is only determined by the spatial part of the transfer
matrix. In section \ref{sec:loop}, we study simple loop scatterers
of arbitrary shapes in the presence of either magnetic field or
the spin-orbit interaction. We derive the transfer matrices for
the loops and discuss the onset of electron localization in the
linear chains of the loops. As examples, we discuss the circular
loop scatterers, the square loop scatterers, and the triangular
loop scatterers in details along with their transfer matrices.
Throughout the paper, we assume that the loops and the leads are
made by single-channel quantum wires for simplicity. The presence
or the absence of electron localization in the loop chains is
shown to be easily known from the criterion functions of the loop
scatterers. In section \ref{sec:probability}, we present a simple
formalism for integrating the transmission probability over $k$.
The averaged transmission probabilities for the loop chains
discussed in the paper are calculated in this section. In section
\ref{sec:conclusion}, we give our conclusion.

\section{scatterers with an input lead and an output lead
\label{sec:scatter}}
\subsection{Transfer matrix for spin-blind scatterers}
\begin{figure}
\hspace*{-1.5cm}
\includegraphics{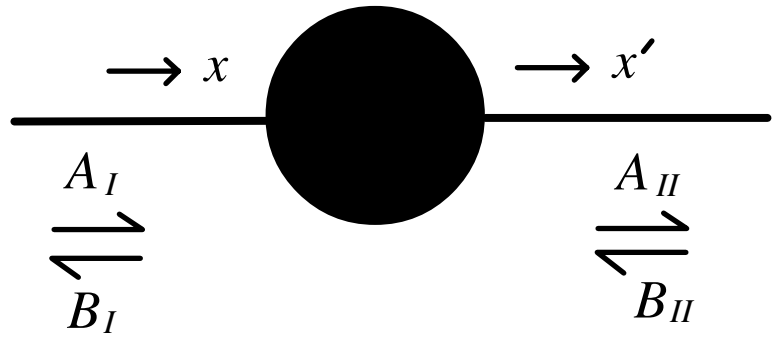}
\vspace{-12cm} \caption{Single-channel scatterer with $A_I,
B_{II}$ and $A_{II}, B_I$ being the amplitudes of the incoming and
the outgoing electronic waves, respectively. The dark circle
represents the scatterer. \label{fig:scatter}}
\end{figure}

In this section we describe the transfer matrix for an arbitrary
spin-blind scatterer with only an input lead and an output lead
connected to it. Consider a scatterer as shown in
Fig.\ref{fig:scatter}. The dark circle represents the unknown
scatterer. Two single-channel quantum wires are connected to the
scatterer such that electrons can move in the wires and tunnel
through the scatterer. By assuming that electrons move as free
particles in the wires, the electron wave functions on the two
leads are written as $\psi_I(x)=A_Ie^{ikx}+B_Ie^{-ikx}$ and
$\psi_{II}(x^\prime)=A_{II}e^{ikx^\prime}+ B_{II}e^{-ikx^\prime}$
, respectively. Here $k$ denotes the injection wave vector for
electrons, $x$ and $x^\prime$ denote the coordinates on the leads,
and the constants $A_I, B_I, A_{II}, B_{II}$ are the amplitudes of
the waves. The conservation of electron current then implies that
$|A_I|^2-|B_I|^2=|A_{II}|^2-|B_{II}|^2$. In general, quantum
mechanics also implies the following linear relation between the
amplitudes
\begin{eqnarray}
\left(
\begin{array}{c}
  A_{II} \\
  B_{II} \\
\end{array}
\right)=T\left(
\begin{array}{c}
  A_I \\
  B_I \\
\end{array}
\right), \label{eqn:transferT}
\end{eqnarray}
\noindent where $T$ is called the transfer matrix for the
scatterer. It is obvious that $T$ is an element of $U(1,1)$ group.
The most general form of $T$ can be parameterized as
\begin{eqnarray}
T=e^{i\alpha}\left(
\begin{array}{cc}
  e^{i\beta}\cosh\theta & e^{-i\gamma}\sinh\theta \\
  e^{i\gamma}\sinh\theta &  e^{-i\beta}\cosh\theta\\
\end{array}
\right),\label{eqn:Tparameters}
\end{eqnarray}
\noindent where $\alpha,\beta,\theta,$ and $\gamma$ are the
introduced parameters to be determined by the nature of the
scatterer and the injection wave vector of electrons.

With the parameterized $T$ given in Eq.(\ref{eqn:Tparameters}), we
find the eigenvalues $\lambda_{\pm}$ and the corresponding
eigenvectors $\vec{\phi}_{\pm}$ as follows:
\begin{eqnarray}
\lambda_{\pm}&=&e^{i\alpha}(\cosh\theta\cos\beta{\pm}\sqrt{(\cosh\theta
\cos\beta)^2-1}) \label{eqn:Teigenval}\\
\vec{\phi}_{\pm}&=&(\cosh\theta\sin\beta\mp i \sqrt{(\cosh\theta
\cos\beta)^2-1}, -i e^{i\gamma} \sinh\theta).
\label{eqn:Teigenvec}
\end{eqnarray}
\noindent The transfer matrix $T$ is then diagonalized with the
help of the $2\times 2$ matrix $\Lambda$
\begin{eqnarray}
T_D\equiv\left(
\begin{array}{cc}
  \lambda_+ & 0 \\
  0 & \lambda_- \\
\end{array}
\right)=\Lambda^{-1}T\Lambda, \label{eqn:Tdiagonal}
\end{eqnarray}
\noindent where the first and the second columns of $\Lambda$ are
the eigenvectors $\vec{\phi}_+$ and $\vec{\phi}_-$, respectively.

\subsection{Chain of identical spin-blind scatterers}
\begin{figure}
\hspace*{-1.5cm}
\includegraphics{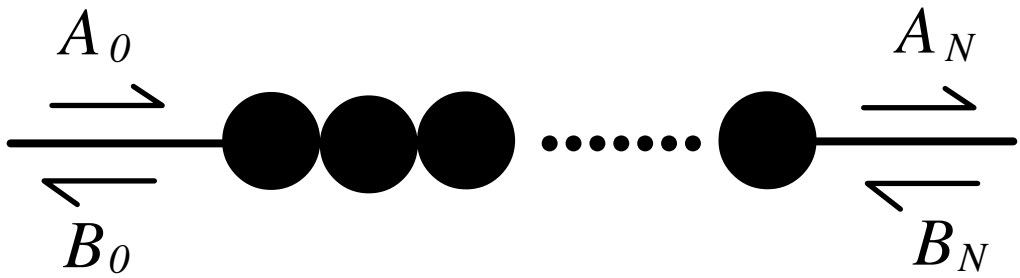}
\vspace{-13cm} \caption{A chain of $N$ identical
scatterers.\label{fig:chain}}
\end{figure}

Consider a linear chain of $N$ identical spin-blind scatterers as
shown in Fig.\ref{fig:chain}. The electronic wave amplitudes at
the two ends of the chain are related by
\begin{eqnarray}
\left(
\begin{array}{c}
  A_N \\
  B_N \\
\end{array}
\right)=\Lambda T^N_D \Lambda^{-1}
\left(
\begin{array}{c}
  A_0 \\
  B_0 \\
\end{array}
\right). \label{eqn:chainEq}
\end{eqnarray}
\noindent Now let $M$ denote the transformation matrix $M\equiv
\Lambda T^N_D \Lambda^{-1}$. Suppose that an electron is injected
from the left end of the chain. The electron will tunnel through
the chain of scatterers with the tunnelling amplitude $t$, or be
reflected by the scatterers with the reflection amplitude $r$,
\begin{eqnarray}
t&=&\frac{\det{M}}{M_{22}}=\frac{e^{i2N\alpha}}{M_{22}},
\label{eqn:transmission} \\
r&=&\frac{-M_{21}}{M_{22}}. \label{eqn:reflection}
\end{eqnarray}
\noindent Here $M_{ij}$ denote the entries of $M$
\begin{eqnarray}
M_{21}&=&\frac{e^{i\gamma}(\lambda^N_+ -\lambda^N_-)\sinh\theta}
{2\sqrt{(\cosh\theta \cos\beta)^2 -1}}, \label{eqn:M21} \\
M_{22}&=&\frac{\lambda^N_+ + \lambda^N_-}{2}-i\frac{(\lambda^N_+ -
\lambda^N_-)\cosh\theta \sin\beta}{2\sqrt{(\cosh\theta
\cos\beta)^2 -1}}. \label{eqn:M22}
\end{eqnarray}

As the number of scatterers $N$ becomes large, the electronic
transmission will be highly suppressed for $(\cosh\theta
\cos\beta)^2>1$. In this low transmission scenario, the
transmission probability is roughly
\begin{eqnarray}
|t|^2\simeq \frac{4[(\cosh\theta \cos\beta)^2-1]}{|\lambda|^{2N}
\sinh^2\theta}, \label{eqn:lowTransmission}
\end{eqnarray}
\noindent where $|\lambda|=\max(|\lambda_+|, |\lambda_-|)$ is the
larger one of the absolute values of the two eigenvalues. The
electronic transmission will finally be turned off as $N$ goes to
infinity. For $(\cosh\theta \cos\beta)^2<1$, the eigenvalues can
be written as $\lambda_\pm = e^{i(\alpha \pm \Omega)} $ with
$\Omega \equiv \cos^{-1}(\cosh\theta \cos\beta)$. In this scenario
the electronic transmission is high and the transmission
probability is found as exactly
\begin{eqnarray}
|t|^2=\frac{\sin^2\Omega}{\sin^2\Omega + \sinh^2\theta
\sin^2(N\Omega)}. \label{eqn:highT}
\end{eqnarray}
\noindent Obviously, the transmission probability oscillates
rapidly in $\Omega$ for large $N$. In the large $N$ limit, the
averaged transmission probability over the small interval
$[\Omega, \Omega+2\pi/N]$ is
\begin{eqnarray}
|t|^2=\frac{|\sin\Omega|}{\sqrt{\sin^2\Omega + \sinh^2\theta}}.
\label{eqn:aveT}
\end{eqnarray}
\noindent Finally, the transmission probability becomes unity when
$(\cosh\theta \cos\beta)^2=1$.

Based upon the above discussion, the value of $(\cosh\theta
\cos\beta)^2$ is critical in determining the characteristics of
the electronic transmission in a chain of spin-blind scatterers.
In real situation, $(\cosh\theta \cos\beta)^2$ is a function of
several dynamical and geometrical factors of the scatterer. In the
remaining parts of the paper, we call the function as the
criterion function for electronic transmission. For examples, the
injection wave vector $k$, the Aharonov-Bohm flux $\Phi_{AB}$, and
the ring radius $a$ together will determine the $(\cosh\theta
\cos\beta)^2$ value of the equal-arm ring scatterer in the
presence of a perpendicular magnetic field\cite{RashbaMagRings}
 \begin{eqnarray}
 (\cosh\theta\cos\beta)^2=\frac{\cos^2(\pi k a)}{\cos^2
 ({\Phi_{AB}/ 2})}. \label{eqn:ringMag}
 \end{eqnarray}
\noindent As long as $ka$ is not a half-integer, the low
electronic transmission condition $(\cosh\theta \cos\beta)^2>1$ is
always held when $\Phi_{AB} = (2n+1)\pi$ for integers $n$.
Extremely low transmission still survives even if we integrate the
transmission probability over $k$. Therefore electron localization
occurs in the chain of equal-arm rings in the presence of a
perpendicular magnetic field.

\subsection{Spin-sensitive scatterers}
In this subsection we briefly discuss spin-sensitive scatterers.
Once again, the scatterer is connected to two leads as shown in
Fig.\ref{fig:scatter}. The wave functions in the leads $I$ and
$II$ are written as $\psi_I(x)=A_I e^{ikx}
\chi_I+B_Ie^{-ikx}\eta_I$ and $\psi_{II}(x^\prime)=A_{II}
e^{ikx^\prime} \chi_{II}+B_{II}e^{-ikx^\prime}\eta_{II}$. Here
$\chi_I, \chi_{II}, \eta_I$ and $\eta_{II}$ are normalized
spinors, $A_I, A_{II}, B_I$ and $B_{II}$ are amplitudes of the
waves. In general, $A_{II}, B_{II}, \chi_{II}$ and $\eta_{II}$ can
be viewed as functions of $A_I, B_I, \chi_{I}$ and $\eta_{I}$. If
we write the electron current fluxes into the spin-up and the
spin-down parts by
$|A_i|^2=|A_i^{\uparrow}|^2+|A_i^{\downarrow}|^2$ and
$|B_i|^2=|B_i^{\uparrow}|^2+|B_i^{\downarrow}|^2$, then the
current conservation gives $|A_I^\uparrow|^2 +
|A_I^\downarrow|^2-|B_I^\uparrow|^2 - |B_I^\downarrow|^2 =
|A_{II}^\uparrow|^2 + |A_{II}^\downarrow|^2-|B_{II}^\uparrow|^2 -
|B_{II}^\downarrow|^2$. As easily seen from the equation, the
transfer matrix for the spin-sensitive scatterer must respect
$U(2,2)$ symmetry.

In some cases, the symmetry for the transfer matrix is smaller.
For example, the symmetry is $T \otimes S$ for the ring-shape and
the diamond-shape scatterers with Rashba spin-orbit
coupling\cite{RashbaMagRings, RashbaNetDetail, RashbaNetwork}.
Here $T$ is an element of $U(1,1)$ group and acts on the spatial
part of the electron wave function as in Eq.(\ref{eqn:transferT}),
while $S$ denotes an element of $U(2)$ group and acts on the spin
part of the electron wave functions, $\chi_{II}=S\chi_{I}$ and
$\eta_{II}=S\eta_{I}$. Obviously, the total electronic
transmission probability is merely determined by $T$ and has
nothing to do with the spin-part symmetry $S$. In these cases, the
spin-sensitive scatterers are effectively spin-blind in discussing
their total transmission probabilities.

\section{quantum wire loops as scatterers\label{sec:loop}}
In this section, we first discuss simple loop scatterers in the
presence of either a perpendicular magnetic field or the
spin-orbit interaction. We derive the transfer matrices for the
loop scatterers. It is found that the transfer matrices merely
depend on the injection wave vector $k$, the arm lengths of the
loops, and the Aharonov-Bohm phases or the spin-orbit coupling
strength. We then apply our results to the linear chains of the
circular loops\cite{RashbaMagRings}, the square
loops\cite{RashbaNetDetail, RashbaNetwork} and the triangular
loops. We take all these quantum wire loops as scatterers and find
their transfer matrices. Given the transfer matrices, we thus
obtain the criterion function $(\cosh\theta \cos\beta)^2$ as well
as the parameter $\sinh\theta$ that appears in the parameterized
transfer matrix in Eq.(\ref{eqn:Tparameters}).

\subsection{Simple loops in the presence of a perpendicular
magnetic field}

\begin{figure}
\hspace*{-1.5cm}
\includegraphics{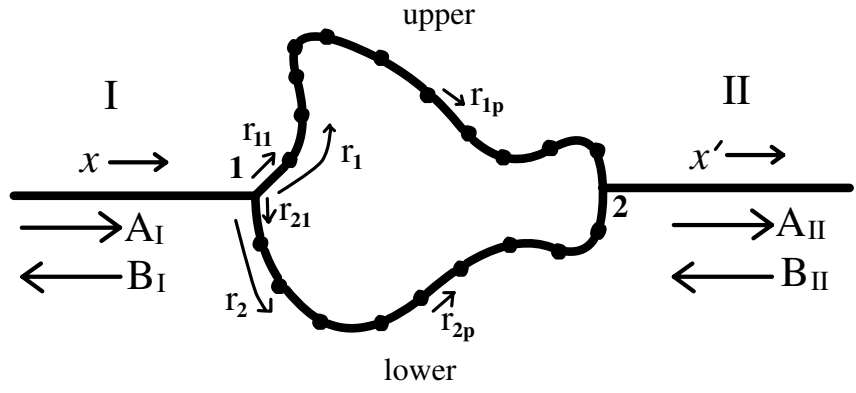}
\vspace{-12cm} \caption{A simple loop with two attached leads.}
\label{fig:genloop}
\end{figure}

Consider the scatterer that is made by a simple quantum wire loop
with two leads in the presence of a perpendicular magnetic field
as shown in Fig.\ref{fig:genloop}. The simple loop can be of any
shapes. Let $r_1$ and $r_2$ denote the path lengths measured from
the node $1$ along the upper arm and the lower arm, respectively.
The total length of the upper arm of the loop is $L_1$ while the
total length of the lower arm is $L_2$. In general, the two arm
lengths may not be equal to each other.

Suppose that an electron is injected into the loop with energy
$\varepsilon = \hbar^2 k^2/2m$, then the electron wave functions
in the two leads and in the loop arms are given by
\begin{eqnarray}
\psi_I(x) &=& A_I e^{ikx} + B_I e^{-ikx}, \label{eqn:function01}\\
\psi_{II}(x) &=& A_{II} e^{ikx^\prime} + B_{II}
e^{-ikx^\prime},\label{eqn:function02}
 \\
\psi_{up}(r_1)&=& e^{i\frac{e}{\hbar} \int_0^{r_1} \vec{A} \cdot
d\vec{r}_1} (a_1 e^{ikr_1} +b_1 e^{-ikr_1}), \\
\psi_{low}(r_2)&=& e^{i\frac{e}{\hbar} \int_0^{r_2} \vec{A} \cdot
d\vec{r}_2} (a_2 e^{ikr_2} +b_2 e^{-ikr_2}),
\end{eqnarray}

\noindent where $k$ is the injection wave vector, $e$ is the
electric charge for electrons, $m$ denotes the effective mass for
electrons , $\vec{A}$ denotes the vector potential for the
magnetic field, and $x$ and $x^\prime$ are the coordinates on the
leads that are measured from the node $1$ and the node $2$,
respectively. Here $\psi_{up}$ is the wave function in the upper
arm and $\psi_{low}$ denotes the wave function in the lower arm.
The coefficients $A_I, B_I, A_{II}, B_{II}$ and $a_i, b_i$ are the
amplitudes of the electronic waves. By considering the continuity
of wave function and the current conservation at the junctions, we
get the transfer matrix for the loop scatterer
\begin{eqnarray}
\left(
\begin{array}{c}
  A_{II} \\
  B_{II} \\
\end{array}
\right) = e^{i(\frac{\Phi_1+ \Phi_2}{2}-\Theta_{AB})} \left(
\begin{array}{cc}
  t_2 & -i t_1 \\
  i t_1 & t_2^* \\
\end{array}
\right) \left(
\begin{array}{c}
  A_I \\
  B_I \\
\end{array}
\right), \label{eqn:genloopT}
\end{eqnarray}
\noindent with
\begin{eqnarray}
\Theta_{AB}&=& \arctan[(\frac{\sin\theta_1 - \sin\theta_2}
{\sin\theta_1+\sin\theta_2})\tan(\frac{\Phi_{AB}}{2})],\label{eqn:ThegenLP}\\
 t_2&=&\frac{4\sin(\theta_1+\theta_2) +i[\cos(\theta_1
-\theta_2)-5\cos(\theta_1+\theta_2)+ 4\cos(\Phi_{AB}) ]}{4
\sqrt{\sin^2\theta_1 +\sin^2\theta_2 +2\sin\theta_1 \sin\theta_2
 \cos(\Phi_{AB})}}, \label{eqn:t2genLP}\\
t_1&=& \frac{\cos(\theta_1-\theta_2) +3\cos(\theta_1 +\theta_2)
-4\cos(\Phi_{AB})}{4\sqrt{\sin^2\theta_1 +\sin^2\theta_2
+2\sin\theta_1 \sin\theta_2 \cos(\Phi_{AB})}}. \label{eqn:t1genLP}
\end{eqnarray}
\noindent Here $t_2^*$ is the complex conjugate of $t_2$, and
$\Phi_1$ and $\Phi_2$ are the Aharonov-Bohm phases acquired by the
electrons travelling in the upper arm and the lower arm,
respectively. $\Phi_{AB}\equiv\Phi_1-\Phi_2$ is the Aharonov-Bohm
flux of the loop, and the phases are defined as $\theta_1\equiv k
L_1$ and $\theta_2\equiv k L_2$. Comparing Eq.(\ref{eqn:genloopT})
to (\ref{eqn:Tparameters}), we find the transfer matrix parameter
$\sinh\theta = t_1$ and the criterion function
\begin{eqnarray}
(\cos\beta \cosh\theta)^2 = \frac{\sin^2(\theta_1 +\theta_2)}{
\sin^2\theta_1 +\sin^2\theta_2 +2\sin\theta_1 \sin\theta_2
\cos(\Phi_{AB})}. \label{eqn:criteriumGenlp}
\end{eqnarray}
\noindent Eq.(\ref{eqn:criteriumGenlp}) shows that the low
electronic transmission condition $(\cos\beta \cosh\theta)^2 > 1$
cannot be always satisfied for all $k$ values unless for
$L_1=L_2$. When $L_1=L_2=L$, the criterion function becomes
$(\cos\beta \cosh\theta)^2 = \cos^2(kL)/\cos^2(\Phi_{AB}/2)$. For
the equal-arm loop chains in the presence of a perpendicular
magnetic field, electron localization will occur at
$\Phi_{AB}=(2n+1)\pi$ for all integers $n$.

\subsection{Simple loop with spin-orbit interaction}
Consider a general loop scatterer with spin-orbit interaction as
shown in Fig.\ref{fig:genloop}. In general, the upper arm of the
loop can be viewed as being made by $N_1$ linking bonds connected
in series. The lower arm is viewed as being made by $N_2$ linking
bonds connected in series. Typically, the numbers of bonds $N_1$
and $N_2$ could be very large. Let ${r}_{1p}$ and ${r}_{2p}$
denote the coordinates of the $p$-th linking bonds for the upper
arm and the lower arm, respectively. The beginning of the $p$-th
linking bonds for the arms are labelled as $r_{1p}=0$ and
$r_{2p}=0$, while the ends of the bonds are labelled as
$r_{1p}=\ell_{1p}$ and $r_{2p}= \ell_{2p}$. The Hamiltonian
$H_{ip}$ ($i=1,2$) for an electron moving in the $p$-th linking
bonds in the upper arm or the lower arm is given by
\begin{eqnarray}
H_{ip}=-\frac{\hbar^2}{2m}[\frac{\partial}{\partial r_{ip}}- i
k_{so} (\vec{\sigma} \cdot(\hat{z} \times
\hat{r}_{ip}))]^2-\frac{\hbar^2 k_{so}^2}{2m}, \hspace{0.5cm}
i=1,2. \label{eqn:Hq}
\end{eqnarray} where $k_{so}$ is the coupling strength for spin-orbit
interaction, $m$ is the effective mass of electron, $\hat{z}$ is
the unit direction normal to the loop, and $\hat{r}_{ip}$ is the
unit direction along the $p$-th linking bonds in the arms. The
wave function $\psi_{up}(r_{1p})$ with the energy $\varepsilon =
\hbar^2 k^2/2m$ in the upper arm is thus obtained by
\begin{eqnarray}
\psi_{up}(r_{1p})&=&e^{ik_{so} r_{1p}(\vec{\sigma} \cdot(\hat{z}
\times \hat{r}_{1p}))}S^{(1)}_{p-1}S^{(1)}_{p-2} \cdots S^{(1)}_1
[a_1 e^{i q r_1} + b_1 e^{-iq r_1}], \label{eqn:psiup}
\end{eqnarray}
\noindent where $a_1$ and $b_1$ are constant spinors, $k$ is the
injection wave vector, $q \equiv \sqrt{k^2+k_{so}^2}$ is the wave
vector for the electrons moving in the arms of the loop,
$r_1=r_{1p}+\sum_{n=1}^{p-1} \ell_{1n}$ is the path length
measured from the node $1$ to the position $r_{1p}$, and the
spin-rotation operators $S^{(1)}_{n}$ are defined as $S^{(1)}_n
\equiv e^{i k_{so} \ell_{1n} (\vec{\sigma} \cdot(\hat{z} \times
\hat{r}_{1n}))}$. Similarly, the wave function in the lower arm of
the loop is given by
\begin{eqnarray}
\psi_{low}(r_{2p})&=&e^{ik_{so} r_{2p}(\vec{\sigma} \cdot(\hat{z}
\times \hat{r}_{2p}))}S^{(2)}_{p-1}S^{(2)}_{p-2} \cdots S^{(2)}_1
[a_2 e^{iq r_2} + b_2 e^{-iq r_2}], \label{eqn:psilow}
\end{eqnarray}
\noindent where $a_2$ and $b_2$ are constant spinors, $r_2=r_{2p}
+ \sum_{n=1}^{p-1} \ell_{2n}$ is the path length measured from the
node $1$ to the position $r_{2p}$, and the spin-rotation operators
$S^{(2)}_{n}$ are defined as $S^{(2)}_n \equiv e^{i k_{so}
\ell_{2n} (\vec{\sigma} \cdot(\hat{z} \times \hat{r}_{2n}))}$. At
the node $2$, the wave functions in the arms become
\begin{eqnarray}
\psi_{up}(r_1=L_1)&=&S_1[a_1 e^{iq
L_1} + b_1 e^{-iq L_1}], \\
\psi_{low}(r_2=L_2)&=& S_2 [a_2 e^{iq L_2} + b_2 e^{-iq L_2}],
\end{eqnarray}
\noindent where $S_1$ is the total spin-rotation operator for the
upper arm and $S_2$ is the total spin-rotation operator for the
lower arm,
\begin{eqnarray}
S_1&\equiv& S^{(1)}_{N_1} S^{(1)}_{N_1 -1} \cdots S^{(1)}_1, \label{eqn:S1}\\
S_2&\equiv& S^{(2)}_{N_2}S^{(2)}_{N_2-1} \cdots S^{(2)}_1.
\label{eqn:S2}
\end{eqnarray}
\noindent Here $L_1$ and $L_2$ denote the arm lengths of the upper
arm and the lower arm, respectively.

On the other hand, the wave functions in the leads are given by
\begin{eqnarray}
\psi_I(x)&=& A_I e^{ikx}+ B_I e^{-ikx}, \label{eqn:psi1SO}\\
\psi_{II}(x^\prime)&=& A_{II} e^{ikx^\prime} + B_{II}
e^{-ikx^\prime} \label{eqn:psi2SO}.
\end{eqnarray}
\noindent Here $x$ is the coordinate on the input lead with $x=0$
being at the node $1$, and $x^\prime$ is the coordinate on the
output lead with $x^\prime=0$ being at the node $2$. The spinor
amplitudes $A_{II}$ and $B_{II}$ are connected to the $A_I$ and
$B_I$ through the continuity of wave function and current
conservation at the nodes $1$ and $2$. The continuity of wave
function at the nodes $1$ and $2$ gives
\begin{eqnarray}
A_I+B_I&=&a_1+b_1=a_2+b_2, \\
A_{II}+B_{II}&=& S_1(a_1 e^{iqL_1}+b_1e^{-iqL_1}) = S_2(a_2
e^{iqL_2}+b_2e^{-iqL_2}). \label{eqn:contWaveSO}
\end{eqnarray}
\noindent Current conservation at the nodes leads to the
Griffith's boundary conditions as follows
\begin{eqnarray}
\frac{k}{q}(A_I-B_I)&=&a_1-b_1+a_2-b_2,\\
\frac{k}{q}(A_{II}-B_{II})&=& S_1(a_1 e^{iqL_1}-b_1e^{-iqL_1}) +
S_2(a_2 e^{iqL_2}-b_2e^{-iqL_2}). \label{eqn:Griffith}
\end{eqnarray}
\noindent Solving the equations in the above, we get the
transformation
\begin{eqnarray}
\left(
\begin{array}{c}
  A_{II} \\
  B_{II} \\
\end{array}
\right)= \left(
\begin{array}{cc}
  t_2 & -it_1 \\
  it_1 & t_2^* \\
\end{array}
\right)\left(
\begin{array}{c}
  S(k_{so},q)A_I \\
  S(k_{so},q)B_I \\
\end{array}
\right). \label{eqn:transferTSO}
\end{eqnarray}
\noindent Here the $2\times 2$ unitary matrix $S(k_{so},q)$ is the
transfer matrix in spin dimension and is defined as
\begin{equation}
S(k_{so},q)=\frac{1}{\sqrt{f(k_{so},q)}}[S_1+
\frac{\sin(\theta_1)}{\sin(\theta_2)}S_2],\label{eqn:SSO}
\end{equation}
\noindent with $\theta_1\equiv qL_1$ and $\theta_2 \equiv qL_2$,
and
\begin{equation}
f(k_{so},q)\equiv 1+\frac{\sin^2(\theta_1)}{\sin^2(\theta_2)}+2
\cos(\Theta_{so})\frac{\sin(\theta_1)}{\sin(\theta_2)},
\label{eqn:fSO}
\end{equation}
\noindent where $\cos(\Theta_{so})\equiv \frac{1}{4}\texttt{Tr}(
S_2S_1^{-1} + S_1S_2^{-1})$ is a function of $k_{so}$. It is noted
that $S_2S_1^{-1}+S_1S_2^{-1}=2\cos(\Theta_{so})\mathbb{I}_2$,
where $\mathbb{I}_2$ denotes the $2\times 2$ identity matrix. The
matrix elements in Eq.(\ref{eqn:transferTSO}) are obtained by
\begin{eqnarray}
t_2&=&\frac{\sin(\theta_1+\theta_2)}{\sin(\theta_2)\sqrt{f(k_{so},q)}}
+ i \{\frac{k^2 \sin^2(\theta_1) + q^2[f(k_{so},q)-
\csc^2(\theta_2) \sin^2(\theta_1 +\theta_2)]} {2 kq \sin(\theta_1)
\sqrt{f(k_{so},q)}}\},
\label{eqn:t2SO}\\
t_1&=& \frac{k^2\sin^2(\theta_1) - q^2[f(k_{so},q)-
\csc^2(\theta_2) \sin^2(\theta_1 +\theta_2)]}{2kq \sin(\theta_1)
\sqrt{f(k_{so},q)}}. \label{eqn:t1SO}
\end{eqnarray}
\noindent

Obviously, the transfer matrix in Eq.(\ref{eqn:transferTSO}) is of
the type $T\otimes S$ and the total electronic transmission is
merely determined by the matrix elements of $T$. The parameter
$\sinh\theta$ in Eq.(\ref{eqn:Tparameters}) is equal to $t_1$ in
Eq.(\ref{eqn:t1SO}), and the criterion function for the chain of
loops is found by
\begin{eqnarray}
(\cosh\theta\cos\beta)^2 =
\frac{\sin^2(\theta_1+\theta_2)}{\sin^2(\theta_2) f(k_{so},q)}.
\label{eqn:CtmSO}
\end{eqnarray}
\noindent In Eq.(\ref{eqn:CtmSO}), the variables $k_{so}$ and $q$
are not separated generically. When $L_1=L_2$ both $S$ and $f$
become functions of $k_{so}$,
$S(k_{so})=\frac{1}{\sqrt{f(k_{so})}}(S_1+S_2)$ and
$f(k_{so})=2(1+\cos(\Theta_{so}) )$, so that $k_{so}$ and $q$ are
separated in Eq.(\ref{eqn:CtmSO}). In the equal-arm case, the
criterion function is always larger than $1$ for some specific
values of $k_{so}$ that satisfy $f(k_{so})=0$, i.e,
$\cos(\Theta_{so})=-1$. Therefore, electron localization occurs
naturally in the equal-arm loop chains with spin-orbit
interaction.

\subsection{Circular Loops with spin-orbit interaction}
\begin{figure}
\hspace*{-1.5cm}
\includegraphics{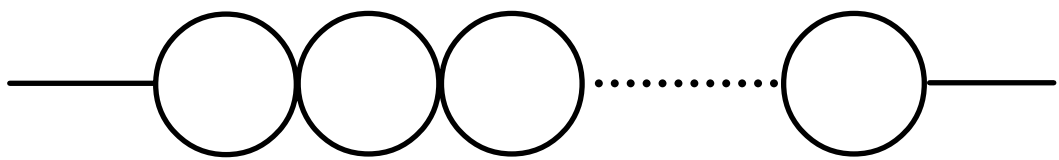}
\vspace{-13cm} \caption{A chain of $N$ identical rings.}
\label{fig:circlechain}
\end{figure}

Assume that an electron with energy $\varepsilon = \hbar^2 k^2/2m$
is injected into the linear chain of circular loops as shown in
Fig.\ref{fig:circlechain}. The radius of the rings is denoted as
$a$. Both the lengths of the upper arm and the lower arm are equal
to each other and are denoted as $L_1=L_2=\pi a$. As discussed in
literature\cite{RashbaMagRings}, the transfer matrix and the
electronic transmission are obtained for the equal-arm ring
scatterer in the presence of the spin-orbit interaction. Here we
derive the transfer matrix again by using the results obtained in
the previous subsections. It is found that electron localization
will occur in the chain of rings with the spin-orbit interaction.

\begin{figure}
\hspace*{-1.5cm}
\includegraphics{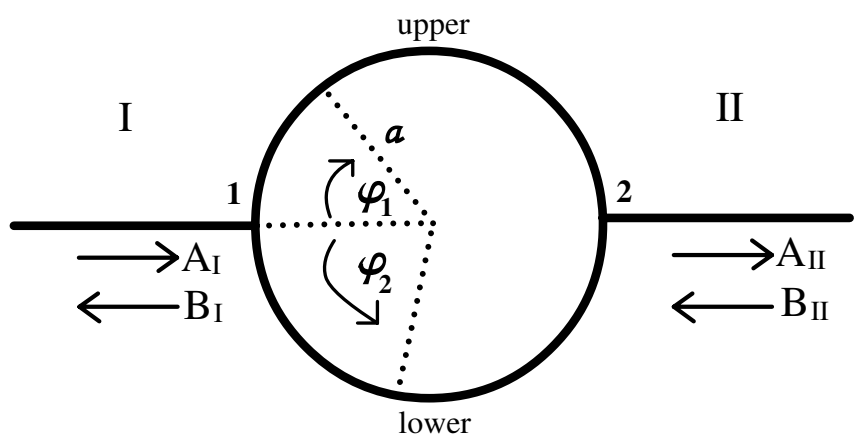}
\vspace{-12cm} \caption{A single ring with two attached leads.}
\label{fig:circle}
\end{figure}

The transfer matrix is given in Eq.(\ref{eqn:transferTSO}). To
know the exact form of the transfer matrix, we need the total
spin-rotation operators $S_1$ and $S_2$ for the upper arms and
lower arms of the ring, respectively. Here we derive the total
spin-rotation operators as follows. Let $\varphi_1$ be the angles
on the upper arm measured from the the node $1$, and $\varphi_2$
denote the angle on the lower arm that is also measured from the
node $1$ as shown in Fig.\ref{fig:circle}. From
Eq.(\ref{eqn:psiup}), the electron wave function $\Psi_{up}$ in
the upper arm is given by $\psi_{up}(\varphi_1)= S_1(\varphi_1)
[a_1 e^{i q a \varphi_1} + b_1 e^{-iq a \varphi_1}]$. We then find
the differential equation for the spin-rotation operator
$S_1(\varphi_1)$ as
\begin{eqnarray}
\frac{dS_1}{d\varphi_1} = i\theta_{so}\sigma_r(\varphi_1)
S_1(\varphi_1) \label{eqn:S1eqn},
\end{eqnarray}
\noindent where $\theta_{so} \equiv k_{so} a$, and $\sigma_r$ is
defined as $\sigma_r(\varphi_1) \equiv [-\cos(\varphi_1)\sigma_1 +
\sin(\varphi_1)\sigma_2]$. By considering that the unitary
operator $S_1(\varphi_1)$ is equal to the identity operator at
$\varphi_1=0$, we obtain $S_1(\varphi_1)$ as
\begin{eqnarray}
S_1(\varphi_1)=\frac{1}{1+\rho^2} \left(
\begin{array}{cc}
  e^{-i\lambda_1 \varphi_1} + \rho^2
  e^{i\lambda_2 \varphi_1} & \rho [e^{-i\lambda_1 \varphi_1}
  - e^{i\lambda_2 \varphi_1}] \\
  \rho[e^{-i\lambda_2 \varphi_1}- e^{i\lambda_1 \varphi_1}]&
   e^{i\lambda_1 \varphi_1} + \rho^2
  e^{-i\lambda_2 \varphi_1} \\
\end{array}
\right), \label{eqn:S1form}
\end{eqnarray}
\noindent where $\rho=({\theta_{so}/ \lambda_2})$,
$\lambda_1=(\sqrt{1+4\theta_{so}^2}-1)/2$, and
$\lambda_2=(\lambda_1+1)$. Similarly, we obtain $S_2(\varphi_2)$
as
\begin{eqnarray}
S_2(\varphi_2)=\frac{1}{1+\rho^2} \left(
\begin{array}{cc}
  e^{i\lambda_1 \varphi_2} + \rho^2
  e^{-i\lambda_2 \varphi_2} & \rho
  [e^{i\lambda_1 \varphi_2}-e^{-i\lambda_2 \varphi_2}] \\
  \rho [e^{i\lambda_2 \varphi_2}-e^{-i\lambda_1 \varphi_2}]&
   e^{-i\lambda_1 \varphi_2} + \rho^2
  e^{i\lambda_2 \varphi_2} \\
\end{array}
\right). \label{eqn:S2form}
\end{eqnarray}
\noindent From Eq.(\ref{eqn:SSO}), we find that $S(k_{so},q)$
becomes a function of $\theta_{so}$ and is given by
\begin{eqnarray}
S(\theta_{so})&=&\frac{1}{2\cos(\lambda_1 \pi)}[S_1(\pi)+S_2(\pi)],
\label{eqn:SforRing} \\
&=& \exp({i \delta \sigma_2}),
\end{eqnarray}
\noindent with $\delta=\arctan(2\theta_{so})$. Compare
Eq.(\ref{eqn:SforRing}) to (\ref{eqn:SSO}), we find that the
function $f(k_{so},q)$ also becomes a function of $\theta_{so}$
and is given by $f(\theta_{so})=4\cos^2({\lambda_1 \pi}) =
4\cos^2(\Phi_{AC}/2)$ with the Aharonov-Casher phase defined as
$\Phi_{AC} \equiv \pi[\sqrt{1+4\theta_{so}^2}-1]$.

Take the above results into Eq.(\ref{eqn:t2SO}) and
(\ref{eqn:t1SO}), we find the transfer matrix for the ring
scatterer
\begin{eqnarray}
\left(
\begin{array}{c}
  A_{II} \\
  B_{II} \\
\end{array}
\right)= \left(
\begin{array}{cc}
  t_2 & -i t_1 \\
  i t_1 & t_2^* \\
\end{array}
\right)\left(
\begin{array}{c}
  S(\theta_{so}) A_I \\
  S(\theta_{so}) B_I \\
\end{array}
\right), \label{eqn:ringOnlySO}
\end{eqnarray}
\noindent with the matrix elements
\begin{eqnarray}
t_2 &=&\frac {4kq\sin(2\pi qa) + i \{k^2[1-\cos(2\pi qa)] +
4q^2[\cos(\Phi_{AC})-\cos(2\pi qa)]\}} {8kq \sin(\pi
qa)\cos(\Phi_{AC}/2)}, \\
t_1&=&\frac {k^2[1-\cos(2\pi qa)] -4q^2[\cos(\Phi_{AC})- \cos(2\pi
qa)]} {8kq \sin(\pi qa)\cos(\Phi_{AC}/2)}. \label{eqn:t1SORing}
\end{eqnarray}
\noindent Comparing Eq.(\ref{eqn:ringOnlySO}) to
Eq.(\ref{eqn:Tparameters}), the parameter $\sinh\theta = t_1$ is
easily read off and the criterion function is found as
\begin{eqnarray}
(\cosh\theta \cos\beta)^2= {\cos^2(\pi qa) \over \cos^2(\Phi_{AC}
/2)}. \label{eqn:ctmSORing}
\end{eqnarray}
\noindent From Eq.(\ref{eqn:ctmSORing}), electron localization
will occur at some specific values of the spin-orbit coupling,
$\theta_{so}=\sqrt{4n^2-1}/2$ for all integers $n\neq 0$, for a
linear chain of the ring scatterers in the large $N$ limit.

\subsection{Circular loops in the presence of a perpendicular
 magnetic field}
 Consider the same circular loop in the presence of a perpendicular
magnetic field $\vec{B}=B\hat{z}$ but without the spin-orbit
interaction. The electron wave functions in the two leads are
given in Eq.(\ref{eqn:function01}) and (\ref{eqn:function02}).
From the results in Eq.(\ref{eqn:genloopT}), (\ref{eqn:ThegenLP}),
(\ref{eqn:t2genLP}), and (\ref{eqn:t1genLP}), we find the transfer
matrix for the ring scatterer
\begin{eqnarray}
\left(
\begin{array}{c}
  A_{II} \\
  B_{II} \\
\end{array}
\right)=\left(
\begin{array}{cc}
  t_2 & -i t_1 \\
  i t_1 & t_2^* \\
\end{array}
\right)\left(
\begin{array}{c}
  A_I \\
  B_I \\
\end{array}
\right),\label{ringOnlyMag}
\end{eqnarray}
\noindent with
\begin{eqnarray}
t_2&=& \frac{4\sin(2\pi ka)+i[1-5\cos(2\pi ka) + 4
\cos(\Phi_{AB})]}{8\sin(\pi ka) \cos(\Phi_{AB}/2)},
\label{eqn:t2circleMag}\\
t_1&=& \frac{1+3\cos(2\pi ka)-4\cos(\Phi_{AB})} {8\sin(\pi
ka)\cos(\Phi_{AB}/2)}\label{eqn:t1circleMag}.
\end{eqnarray}
\noindent Here $\Phi_{AB}\equiv e\pi a^2 B/\hbar$ denotes the
Aharonov-Bohm flux in the ring. From Eq.(\ref{eqn:t1circleMag})
and (\ref{eqn:t2circleMag}), we read off the parameter $\sinh
\theta = t_1$ and the criterion function for the chain of rings is
given in Eq.(\ref{eqn:ringMag}). From Eq.(\ref{eqn:ringMag}) we
know that electron localization will occur for $\Phi_{AB} =
(2n+1)\pi$ for all integers $n$.

\subsection{square loops with spin-orbit interaction}
\begin{figure}
\hspace*{-2cm}
\includegraphics{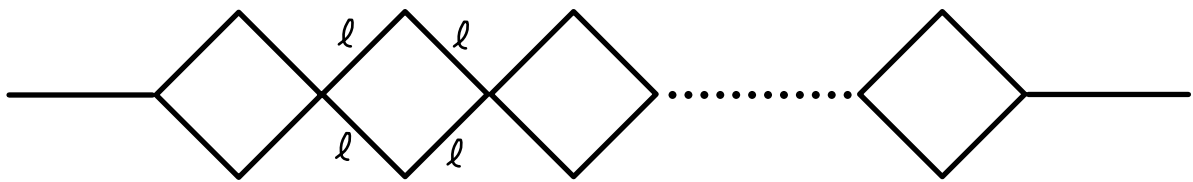}
\vspace{-14cm} \caption{A chain of $N$ identical square loops.}
\label{fig:squarechain}
\end{figure}

Consider that an electron with energy $\varepsilon = \hbar^2
k^2/2m$ is injected to the linear chain of square loops as shown
in Fig.\ref{fig:squarechain}. The length of each side of the
squares is denoted as $\ell$. Therefore both the lengths of the
upper arm and the lower arm are equal $L_1=L_2=2\ell$. As
discussed in literature\cite{RashbaNetwork}, electron localization
will occur in the square loop chain with the spin-orbit
interaction. In this paper, we obtain the same results by using
the formulas given in the previous subsections.

From Eq.(\ref{eqn:S1}) and (\ref{eqn:S2}), we find the total
spin-rotation operators for the upper and the lower arms of a
square loop to be
\begin{eqnarray}
S_1&=&e^{i\theta_{so}\sigma_+} e^{i\theta_{so}\sigma_{-}},
\label{eqn:S1square} \\
S_2&=&e^{i\theta_{so}\sigma_-} e^{i\theta_{so}\sigma_{+}},
\label{eqn:S2square}
\end{eqnarray}
\noindent where $\sigma_{\pm} \equiv \frac{1}{\sqrt{2}}(\sigma_2
\pm \sigma_1)$, and $\theta_{so}$ is defined as
$\theta_{so}=k_{so} \ell$. The function $f(k_{so},q)$ in
Eq.(\ref{eqn:fSO}) thus becomes a function of $\theta_{so}$,
$f(\theta_{so})= 4\cos^2\theta_{so}(1+\sin^2\theta_{so})$.

From Eq.(\ref{eqn:transferTSO}), we find the transfer matrix for
the square loop
\begin{eqnarray}
\left(
\begin{array}{c}
  A_{II} \\
  B_{II} \\
\end{array}
\right)=\left(
\begin{array}{cc}
 t_2 & -i t_1 \\
  i t_1 & t_2^* \\
\end{array}
\right)\left(
\begin{array}{c}
S(\theta_{so})A_I \\
S(\theta_{so})B_I \\
\end{array}
\right),\label{eqn:Tsquare}
\end{eqnarray}
\noindent where $t_2$ and $t_1$ are obtained from
Eq.(\ref{eqn:t2SO}) and (\ref{eqn:t1SO}) by taking
$\theta_1=\theta_2=2q \ell\equiv 2\theta_q$
\begin{eqnarray}
t_2&=&\frac{2\cos(2\theta_q)}{\sqrt{f(\theta_{so})}}+ i\{
\frac{k^2\sin^2(2\theta_q) + q^2[f(\theta_{so})-
4\cos^2(2\theta_q)]}{2kq\sin(2\theta_q)\sqrt{f(\theta_{so})}}\}, \\
t_1&=&\frac{k^2\sin^2(2\theta_q)-q^2[f(\theta_{so})-4\cos^2(2\theta_q)]}
{2kq \sin(2\theta_q)\sqrt{f(\theta_{so})}},\label{eqn:t1squareSO}
\end{eqnarray}
\noindent with $q=\sqrt{k^2+k_{so}^2}$. The matrix
$S(\theta_{so})$ is obtained from Eq.(\ref{eqn:SSO})
\begin{eqnarray}
S(\theta_{so})\equiv \frac{1}{\sqrt{f(\theta_{so})}}\left(
\begin{array}{cc}
  2\cos^2\theta_{so} & \sqrt{2}\sin(2\theta_{so}) \\
  -\sqrt{2}\sin(2\theta_{so}) & 2\cos^2\theta_{so} \\
\end{array}
\right).
\end{eqnarray}
\noindent It is noted that $S(\theta_{so})$ is unitary. Once
again, the transfer matrix in Eq.(\ref{eqn:Tsquare}) is of the
type $T\otimes S$. The parameter $\sinh\theta$ in
Eq.(\ref{eqn:Tparameters}) is then easily read off by
$\sinh\theta=t_1$, and the criterion function for the square loop
scatterer is found by
\begin{eqnarray}
(\cosh\theta\cos\beta)^2=\frac{4\cos^2(2\theta_q)}{f(\theta_{so})}.
\label{eqn:criteriumSquare}
\end{eqnarray}
\noindent In this case, electron localization occurs at
$f(\theta_{so})=0$, i.e., $k_{so}\ell = (n+1/2)\pi$ for all
integers $n$.

\subsection{square loops in the presence of a magnetic field}

Consider the same square loop in the presence of a perpendicular
magnetic field $\vec{B}=B\hat{z}$ but without the spin-orbit
interaction. The electron wave functions in the two leads are
given in Eq.(\ref{eqn:function01}) and (\ref{eqn:function02}).
From the results in Eq.(\ref{eqn:genloopT}), (\ref{eqn:ThegenLP}),
(\ref{eqn:t2genLP}), and (\ref{eqn:t1genLP}), we find the transfer
matrix for the square loop scatterer
\begin{eqnarray}
\left(
\begin{array}{c}
  A_{II} \\
  B_{II} \\
\end{array}
\right)=\left(
\begin{array}{cc}
  t_2 & -i t_1 \\
  i t_1 & t_2^* \\
\end{array}
\right)\left(
\begin{array}{c}
  A_I \\
  B_I \\
\end{array}
\right),\label{eqn:TsquareMag}
\end{eqnarray}
\noindent with
\begin{eqnarray}
t_2&=& \frac{4\sin(4\theta_k)+i[1-5\cos(4\theta_k) + 4
\cos(\Phi_{AB})]}{8\sin(2\theta_k) \cos(\Phi_{AB}/2)},
\label{eqn:T11}\\
t_1&=& \frac{1+3\cos(4\theta_k)-4\cos(\Phi_{AB})}
{8\sin(2\theta_k)\cos(\Phi_{AB}/2)}\label{eqn:T12}.
\end{eqnarray}
\noindent Here $\theta_k\equiv k\ell$, and $\Phi_{AB}\equiv
eB\ell^2/\hbar$ denotes the Aharonov-Bohm flux in the square loop.
From Eq.(\ref{eqn:T11}) and (\ref{eqn:T12}), we read off the
parameter $\sinh \theta = t_1$ and the criterion function for the
square loop chain
\begin{eqnarray}
(\cosh\theta\cos\beta)^2=\frac{\cos^2(2\theta_k)}
{\cos^2(\Phi_{AB}/2)}. \label{eqn:squareMag}
\end{eqnarray}
\noindent Obviously, electron localization will occur for
$\Phi_{AB} = (2n+1)\pi$ for all integers $n$.

\subsection{triangular loops in the presence of a magnetic field}
\begin{figure}
\hspace*{-1cm}
\includegraphics{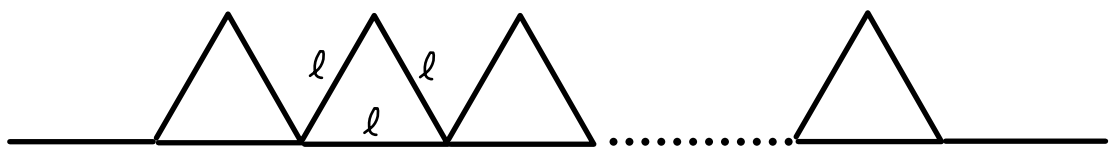}
\vspace{-14cm} \caption{A chain of $N$ identical triangular
loops.} \label{fig:trianglechain}
\end{figure}

In this subsection, we discuss the effect of unequal arm lengths
by considering the triangular loop chain in the presence of a
perpendicular magnetic field $B$ as shown in
Fig.\ref{fig:trianglechain}. Each side of the triangle has the
same length $\ell$. Therefore the length of the upper arm is twice
the length of the lower arm, $L_1=2L_2=2\ell$. Similar to the
square loop chain discussed in the paper, the triangular loop
chain also possesses bipartite structure containing nodes with
different coordination numbers. The lattice nodes have either
coordination number $2$ or coordination number $4$. However,
unlike the square loop chain, the triangular loop chain has no
electron localization even in the presence of a perpendicular
magnetic field.

Once again, the wave functions in the two leads are given in
Eq.(\ref{eqn:function01}) and (\ref{eqn:function02}). By defining
the Aharonov-Bohm flux $\Phi_{AB}=\sqrt{3}eB\ell^2 / 4\hbar$ and
the phase $\theta_k=k\ell$, we obtain the transfer matrix for the
triangular loop from Eq.(\ref{eqn:genloopT}),
(\ref{eqn:ThegenLP}), (\ref{eqn:t2genLP}) and (\ref{eqn:t1genLP})
by
\begin{eqnarray}
\left(
\begin{array}{c}
  A_{II} \\
  B_{II} \\
\end{array}
\right)=e^{i\alpha} \left(
\begin{array}{cc}
  t_2 & -i t_1 \\
  i t_1 & t_2^* \\
\end{array}
\right)\left(
\begin{array}{c}
  A_I \\
  B_I \\
\end{array}
\right),\label{eqn:TtriangleMag}
\end{eqnarray}
\noindent where the phase $\alpha$ is a function of $\Phi_{AB}$
and $\theta_k$, $\alpha = \arctan[\sin\Phi_{AB}/(2\cos\theta_k
+\cos\Phi_{AB})]$. The parameters $t_1$ and $t_2$ are given by
\begin{eqnarray}
t_2&=&\frac{4\sin(3\theta_k)+ i[\cos\theta_k -5\cos(3\theta_k) +
4\cos\Phi_{AB}]}{4\sin\theta_k \sqrt{1+4\cos^2\theta_k+
4\cos\theta_k\cos\Phi_{AB}}}, \label{eqn:t2Tri}\\
t_1&=&\frac{\cos\theta_k +3\cos(3\theta_k)-4\cos\Phi_{AB}}
{4\sin\theta_k \sqrt{1+4\cos^2\theta_k+4\cos\theta_k
\cos\Phi_{AB}}}. \label{eqn:t1Tri}
\end{eqnarray}
\noindent We then read off the parameter $\sinh\theta=t_1$ and the
criterion function for a chain of $N$ identical triangular loops
in the large $N$ limit from Eq.(\ref{eqn:t2Tri}) and
(\ref{eqn:t1Tri}).
\begin{eqnarray}
(\cos\beta \cosh\theta)^2 = \frac{(4\cos^2\theta_k
-1)^2}{1+4\cos^2\theta_k+ 4\cos\theta_k\cos\Phi_{AB}}.
\label{eqn:triangleCriterium}
\end{eqnarray}
\noindent From the criterion function, we know that no electron
localization will occur since no values of $\Phi_{AB}$ could
always satisfy the condition that $(\cos\beta \cosh\theta)^2
> 1$ for all $\theta_k$ values.

\subsection{triangular loops with spin-orbit interaction}
The last example studied is the same triangular loop chain with
only the spin-orbit interaction. The electrons are injected into
the chain with energy $\varepsilon = \hbar^2 k^2/2m$. From
Eq.(\ref{eqn:S1}) and (\ref{eqn:S2}), the total spin-rotation
operators for the upper arm and the lower arm are
\begin{eqnarray}
S_1&=& e^{i\theta_{so}\sigma_{23}}e^{i\theta_{so}\sigma_{12}},
\label{eqn:S1triangle} \\
S_2&=& e^{i\theta_{so}\sigma_2}. \label{eqn:S2triangle}
\end{eqnarray}
\noindent Here $\sigma_{12}$ and $\sigma_{32}$ are defined as
$\sigma_{12}\equiv ( -\frac{\sqrt{3}}{2}\sigma_1 +
\frac{1}{2}\sigma_2)$ and $\sigma_{23}\equiv (
\frac{\sqrt{3}}{2}\sigma_1 + \frac{1}{2}\sigma_2)$, and
$\theta_{so}\equiv k_{so}\ell$.

From Eq.(\ref{eqn:transferTSO}), we obtain the transfer matrix for
the triangular loop scatterer
\begin{eqnarray}
\left(
\begin{array}{c}
  A_{II} \\
  B_{II} \\
\end{array}
\right) = \left(
\begin{array}{cc}
  t_2 & -i t_1 \\
  i t_1 & t_2^* \\
\end{array}
\right)\left(
\begin{array}{c}
  S(k_{so},q) A_I \\
  S(k_{so},q) B_I \\
\end{array}
\right), \label{eqn:triLoopSO}
\end{eqnarray}
\noindent where $t_1$ and $t_2$ are obtained from
Eq.(\ref{eqn:t2SO}) and (\ref{eqn:t1SO}) as
\begin{eqnarray}
t_2 &=& \frac{3-4\sin^2 \theta_q}{\sqrt{f(k_{so},q)}}+ i\{
\frac{k^2\sin^2(2\theta_q) +q^2
[f(k_{so},q)-(3-4\sin^2\theta_q)^2]}{2
kq \sin(2\theta_q) \sqrt{f(k_{so},q)}}\},\label{eqn:t2TriSO}\\
t_1&=& \frac{k^2\sin^2(2\theta_q) -q^2
[f(k_{so},q)-(3-4\sin^2\theta_q)^2]}{2kq \sin(2\theta_q)
\sqrt{f(k_{so},q)}}, \label{eqn:t1TriSO}
\end{eqnarray}
\noindent where $q=\sqrt{k^2+k_{so}^2}$ and $\theta_q \equiv
q\ell$. Here $f(k_{so},q)$ is obtained from Eq.(\ref{eqn:fSO}) by
$f(k_{so},q) = 1+4\cos^2\theta_q +2\cos\theta_q (3\cos\theta_{so}
-\cos^3\theta_{so})$. It is noted that $f(k_{so},q)\geq 0$ is
always held for all $\theta_{so}$ and $\theta_q$. The unitary
matrix $S(k_{so},q)$ is acting on the spinor amplitudes $(A_I,
B_I)$ and is defined as $S(k_{so},q) \equiv (S_1 +2\cos\theta_q
S_2)/\sqrt{f(k_{so},q)}$.

Eq.(\ref{eqn:triLoopSO}) shows that the transfer matrix for the
triangular loop scatterer is of the type $T\otimes S$. Therefore,
the conductance of the loop chain is merely determined by $T$, the
spatial part of the transfer matrix. From
Eq.(\ref{eqn:triLoopSO}), we easily read off the parameter
$\sinh\theta = t_1$ and the criterion function
\begin{eqnarray}
(\cos\beta \cosh\theta)^2 =
\frac{(3-4\sin^2\theta_q)^2}{f(k_{so},q) }.\label{eqn:ctmTriSO}
\end{eqnarray}
\noindent Obviously, electron localization will not occur in the
triangular loop chain since the zero-transmission condition
$(\cos\beta \cosh\theta)^2> 1$ will not hold for all $k$ values
with any specific value of $\theta_{so}$.

\section{Averaged transmission probabilities}\label{sec:probability}

In this section we present the method of integrating the
transmission probability over the injection wave vector $k$ for
the linear loop chains discussed in the paper. Throughout the
paper, the averaged transmission probability will be indicated by
$\langle P \rangle_k$. It is known that finite temperature or
finite voltage will introduce an average over $k$ in a natural
way\cite{RashbaNetwork,RashbaNetDetail}.

To integrate the transmission probability over $k$ we need not
only the knowledge of the criterion function but also the
parameter $\sinh\theta$ as indicated in section \ref{sec:loop}. We
have to find the criterion function and $\sinh\theta$ as functions
of the wave vector $k$ and other physical factors. The expression
for the integrated transmission probability is thus derived in the
integral form. In the numerical calculation, we take the
approximation $q \approx k$. It can be understood as follows.
Taking for the Fermi energy of the single-channel wires $10$ meV,
the loop radius $a$ or the arm length $L \sim a\sim 0.25$ $\mu$m,
and $m/m_e=0.042$ for the effective mass of InAs, yields $k_F a
\sim k_F L \sim 26$. Therefore we could replace $qa$ by $ka$ in
the criterion function in the situation that $k_{so}a \sim k_{so}L
\sim \mathcal{O}(1)$.

\begin{figure}
\includegraphics{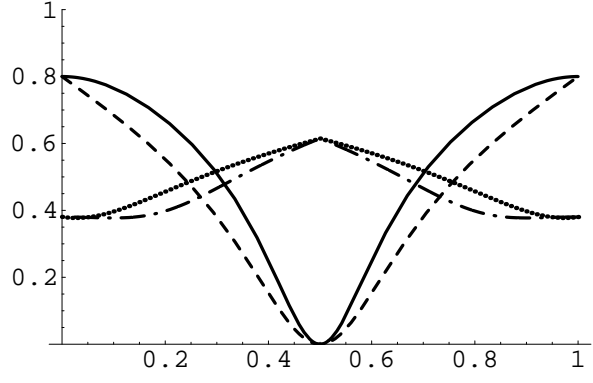}
\caption{The integrated transmission probability $\langle
P\rangle_k$ over the injection wave vector $k$ for the linear
chains of $N$ identical loop scatterers in the large $N$ limit.
Solid line represents $\langle P\rangle_k$ as a function of
$(k_{so}\ell /\pi)$ in the presence of the spin-orbit interaction
for the square loop chain. Doted line and dot-dashed line
represent $\langle P\rangle_k$ as functions of $\Phi_{AB}/\pi$ and
$k_{so}\ell/\pi$ for the triangular loop chain in the presence of
either a perpendicular magnetic field or the spin-orbit effect,
respectively. Dashed line represents $\langle P\rangle_k$ for (a)
the square loop chain in the presence of a perpendicular magnetic
field as a function of $(\Phi_{AB}/\pi)$, and for (b) the circular
loop chain in the presence of either a perpendicular magnetic
field or the spin-orbit interaction as the function of
$(\Phi_{AB}/\pi)$ or $(\Phi_{AC}/\pi)$, respectively. }
\label{fig:ATP}
\end{figure}

In this section we calculate the averaged transmission
probabilities for the loop chains discussed in section
\ref{sec:loop}. As shown in Fig.\ref{fig:ATP}, we find that the
plots of the averaged transmission probabilities resemble each
other for the square and the circular loop chains. Though the
shapes of the loops as well as the physical effects present in the
chains may be different from each other, the square chains and the
circular chains do have something in common. They all have their
loops be annexed at the nodes that equally divide the loops into
upper arms and lower arms with equal arm lengths. The last linear
chain discussed in the section is the triangular loop chain as
shown in Fig.\ref{fig:trianglechain}. The length of the upper arm
is twice the length of the lower arm in the triangular loop. The
plots of the averaged transmission probability are quite different
for the chain, as compared to the plots for the square and the
circular loop chains. Besides, no electron localization will occur
in the chain. The results suggest that electron localization may
not have much to do with the bipartite structure of the lattice
nodes containing different coordination numbers. Instead, it has
strong ties with the arm lengths of the loop scatterers. As
suggested by the results, electron localization is more likely to
occur in the linear loop chains that have equal arms in the loops.

\subsection{square loop chain with spin-orbit effects}
The first example is a linear chain of $N$ square loops in the the
presence of the spin-orbit interaction in the large $N$ limit. In
the numerical calculation, we replace the wave vector $q$ with the
injection wave vector $k$. From Eq.(\ref{eqn:criteriumSquare}),
the criterion function is $(\cosh\theta\cos\beta)^2 =
4\cos^2(2\theta_k)/f(\theta_{so})$. Non-vanishing electronic
transmission occurs at $(\cosh\theta\cos\beta)^2 \leq 1$.
Therefore for a given $\theta_{so}$, non-vanishing transmission
occurs at $2M\pi+\theta_0 \leq 2\theta_k \leq (2M+1)\pi-\theta_0$
with $\theta_0 \equiv \arccos[\sqrt{1-\sin^4\theta_{so}}]$ in the
range $2\theta_k \in [2M\pi, (2M+1)\pi]$ for integer $M$. In the
non-vanishing transmission scenario, we define the variable
$\Omega$ by $\cos\Omega \equiv \cos(2\theta_k) /\cos\theta_0$.
From Eq.(\ref{eqn:aveT}), the averaged transmission probability
$\langle P\rangle_k$ as a function of $\theta_{so}$ is thus given
by integrating $|t^2|$ over $2\theta_k$
\begin{eqnarray}
\langle P \rangle_k=\frac{1}{\pi} \int^{\pi-\theta_0}_
{\theta_0}\frac{\sin\Omega}{\sqrt{\sin^2\Omega+ \sinh^2\theta}}
d(2\theta_k).
\end{eqnarray}
\noindent The parameter $\sinh^2\theta$ is obtained from
Eq.(\ref{eqn:t1squareSO}) as
\begin{eqnarray}
\sinh^2 \theta = \frac{[1+\cos^2\theta_0(3\cos^2\Omega
-4)]^2}{16\cos^2\theta_0 (1-\cos^2\theta_0 \cos^2\Omega)}.
\label{eqn:sinh2}
\end{eqnarray}
\noindent Finally, we obtain the integrated transmission
probability in the integral form
\begin{eqnarray}
\langle P \rangle_k= \frac{\cos\theta_0}{\pi} \int^\pi_0
\frac{\sin^2 \Omega d\Omega}{\sqrt{1-\cos^2\theta_0\cos^2\Omega}
\sqrt{\sin^2\Omega + \sinh^2\theta}}. \label{eqn:aT}
\end{eqnarray}
\noindent Obviously, $\langle P \rangle_k$ is invariant under the
transformations $\theta_{so} \rightarrow \theta_{so} +\pi$ and
$\theta_{so} \rightarrow \pi-\theta_{so}$. The averaged
transmission probability as a function of $(\theta_{so}/\pi)$ is
plotted in Fig.\ref{fig:ATP}. $\langle P\rangle_k$ has its maximum
value at $(\theta_{so}/\pi)=0$ and decreases as
$(\theta_{so}/\pi)$ approaching $1/2$. Electron localization
occurs at $(\theta_{so}/\pi) = 1/2$, consistent with the
prediction derived from the criterion function.

\subsection{square loop chain in the presence of a perpendicular
magnetic field} The second example is the same linear chain of
square loops in the presence of a perpendicular magnetic field
$B$. No spin-orbit interaction is assumed to present in the loops.
For this example, the criterion function for electron transmission
is given in Eq.(\ref{eqn:squareMag}). The non-vanishing
transmission scenarios occur at $\pi-\theta_0 \geq 2\theta_k \geq
\theta_0$, with $\theta_0 \equiv \arccos[|\cos(\Phi_{AB}/2)|]$. By
defining $\cos\Omega= \cos(2\theta_k)/\cos\theta_0$, we find the
parameter $\sinh^2\theta$ has the same expression as that in
Eq.(\ref{eqn:sinh2}). The integrated transmission probability
$\langle P\rangle_k$ is also found to have the same expression as
given in Eq.(\ref{eqn:aT}). The averaged transmission probability
as a function of $(\Phi_{AB}/\pi)$ is also plotted in
Fig.\ref{fig:ATP}. As shown in the figure, the plot is a bit
different to the plot given in the previous example. $\langle
P\rangle_k$ also has its maximum value at $(\Phi_{AB}/\pi)=0$, and
decreases as $(\Phi_{AB}/\pi)$ approaching $1/2$. In this case,
electron localization occurs at $(\Phi_{AB}/\pi) =1/2$.

\subsection{circular loop chain in the presence of a
perpendicular magnetic field}

The third example is the chain of circular loops in the presence
of a perpendicular magnetic field as described in section
\ref{sec:loop}. The criterion function for the chain is given in
Eq.(\ref{eqn:ringMag}) and the parameter $\sinh\theta$ appeared in
the transfer matrix $T$ is equal to the parameter $t_1$ in
Eq.(\ref{eqn:t1circleMag}). Both the criterion function and
$\sinh^2\theta$ are periodic function of $k$.

From Eq.(\ref{eqn:ringMag}), we find the criterion function is no
larger than $1$ for $(1-\theta_o/\pi)\geq ka \geq \theta_0/\pi$ in
the range $ka\in [0,1]$ with $\theta_o \equiv
\arccos(|\cos{(\Phi_{AB}/2)}|)$. By defining $\cos\Omega\equiv
\cos(\pi ka)/\cos(\theta_0)$, the expression of the parameter
$\sinh^2\theta$ is also given in Eq.(\ref{eqn:sinh2}). The
integrated transmission probability $\langle P\rangle_k$ is found
to have the same expression as given in Eq.(\ref{eqn:aT}). The
averaged transmission probability as a function of
$(\Phi_{AB}/\pi)$ is also plotted in Fig.\ref{fig:ATP}. In fact,
the plot is identical to the plot in the previous example.

\subsection{circular loop chain with spin-orbit effects}
The forth example is the chain of circular loops with the
spin-orbit interaction in the absence of the magnetic field. The
criterion function as well we the parameter $\sinh\theta$ are
given in Eq.(\ref{eqn:ctmSORing}) and (\ref{eqn:t1SORing}),
respectively. Here we replace $qa$ with $ka$ in both the criterion
function and in the expression of $t_1$. Thus we find the
criterion function is no larger than $1$ for $2M\pi+\theta_0 \leq
\pi ka \leq (2M+1)\pi-\theta_0$ for integer $M$ with $\theta_0
\equiv \arccos[|\cos(\Phi_{AC}/2)|]$. By defining $\cos\Omega =
\cos(\pi k a)/\cos(\theta_0)$, we find that the parameter
$\sinh^2\theta$ is also given in Eq.(\ref{eqn:sinh2}) and the
integrated transmission probability $\langle P\rangle_k$ is also
found to have the same expression as given in Eq.(\ref{eqn:aT}).
The averaged transmission probability as a function of
$(\Phi_{AC}/\pi)$ is plotted in Fig.\ref{fig:ATP}. Once again, the
plot for the averaged transmission probability is identical to the
plots in the previous two examples.

\subsection{triangular loop chain in the presence of a
perpendicular magnetic field}

The fifth example is the chain of triangular loops in the presence
of a perpendicular magnetic field as described in section
\ref{sec:loop}. The criterion function $(\cos\beta \cosh\theta)^2$
as well as the parameter $\sinh\theta$ can be found in
Eq.(\ref{eqn:triangleCriterium}) and (\ref{eqn:t1Tri}),
respectively. Obviously, both the criterion function and
$\sinh\theta$ are periodic functions of both the wave vector $k$
and the Aharonov-Bohm flux $\Phi_{AB}$. Therefore we will
integrate the transmission probability over $\theta_k$ in the
range $\theta_k \in [0, \pi]$,
\begin{eqnarray}
\langle P\rangle_k = \frac{1}{\pi} \int^{\pi}_{0}
\Theta[1-\cos^2\beta \cosh^2\theta] \{\frac{1- \cos^2\beta
\cosh^2\theta}{1-\cos^2\beta \cosh^2\theta +\sinh^2\theta}\}^{1/2}
d\theta_k. \label{eqn:aTformula}
\end{eqnarray}
\noindent Here $\Theta[1-\cos^2\beta \cosh^2\theta]$ denotes the
step function that equals $1$ for $1 \geq \cos^2\beta
\cosh^2\theta$ and vanishes otherwise. The averaged transmission
probability as a function of $\Phi_{AB}/\pi$ is also plotted in
the range $(\Phi_{AB}/\pi) \in [0,1]$ as shown in
Fig.\ref{fig:ATP}. Different to the previous four examples in this
section, there is no electron localization in the chain of
triangular loops. Besides, $\langle P\rangle_k$ increases as
$(\Phi_{AB}/\pi)$ approaching $1/2$. Once again, the absence of
electron localization is also seen from the criterion function in
Eq.(\ref{eqn:triangleCriterium}).

\subsection{triangular loop chain with spin-orbit interaction}

The last example that we consider is the same triangular loop
chain but in the presence of only the spin-orbit effect. Once
again, we replace $q$ by $k$ in the numerical calculation thus
$\theta_q=\theta_k$. The criterion function and the parameter
$\sinh\theta$ are given in Eq.(\ref{eqn:ctmTriSO}) and
(\ref{eqn:t1TriSO}), respectively. Similarly, they are periodic
functions of $\theta_k$ and $\theta_{so}$. Therefore we will also
integrate the transmission probability over $\theta_k$ in the
range $\theta_k \in [0, \pi]$. The integral formula for the
averaged transmission probability $\langle P \rangle_k$ is also
given in Eq.(\ref{eqn:aTformula}). Obviously, $\langle P
\rangle_k$ is now a function of $\theta_{so}$. $\langle P
\rangle_k$ as a function of $\theta_{so}/\pi$ is also plotted in
Fig.\ref{fig:ATP}. Once again, the plot shows no electron
localization in the chain. This conclusion can also be easily
drawn from the criterion function given in
Eq.(\ref{eqn:ctmTriSO}). As shown in the figure, the plot of
$\langle P \rangle_k$ in this case is similar to the plot for the
same triangular loop chain in the presence of a perpendicular
magnetic field. In fact, the plots of $\langle P \rangle_k$ for
the square and the circular loop chains also resemble each other
even though the chains may have different effects such as the
spin-orbit effect or the Aharonov-Bohm effect. This observation is
quite interesting and seems to be generic for various kinds of the
loop chains. Both the Aharonov-Bohm effect and the spin-orbit
effect may lead to similar behaviors in the averaged conductances
for the linear chains of similar loop structures. Here we think of
the square loop chain and the circular loop chain discussed in the
paper to have similar loop structure, since the annex nodes of a
loop in the chains equally divide the loop into two arms of equal
lengths.

\section{Conclusion}\label{sec:conclusion}

In this paper we discussed the integrated transmission probability
over the wave vector $k$ for electrons moving in the linear chains
of identical loop scatterers, in the presence of either the
spin-orbit interaction or a perpendicular magnetic field. We
showed that the averaged transmission probability is only
determined by some parameters of the transfer matrix of a single
loop scatterer, and can be easily calculated from the knowledge of
the parameters. In fact, the presence or the absence of electron
localization in a linear chain of identical scatterers is easily
known from the square of the real part of one of the transfer
matrix elements, called the criterion function in this paper, of a
single loop scatterer.

The results also show that electron localization is due to the
interplay of the loop geometry and the physical effects such as
the applied magnetic field or the spin-orbit interaction. The loop
geometry plays an important role in the localization phenomenon.
For example, electron localization is found to present in the
linear chains of identical loops with equal arms. This is due to
the fact that totally destructive interference of electronic waves
is more likely to happen in the equal-arm loops. It is understood
in the following ways. Consider an equal-arm loop scatterer in the
presence of a perpendicular magnetic field. The arm length of the
loop is denoted as $L$. An electron will acquire an Aharonov-Bohm
phase $\Phi_1$ along with a phase $kL$ when travelling in the
upper arm of the loop. The electron will also acquire an
Aharonov-Bohm phase $\Phi_2$ along with the same phase $kL$ when
travelling in the lower arm of the loop. Total destruction of
electronic waves at the output lead thus occurs when $e^{i
(kL+\Phi_1)} + e^{i (kL+\Phi_2)} =0$ is satisfied, or equivalently
for $\Phi_{AB}\equiv \Phi_1-\Phi_2=(2n+1)\pi$ no matter what the
value of $k$ is. The total destruction of waves still survives
even after integrating out the wave vector $k$. Therefore electron
localization will occur at the specific values of $\Phi_{AB}$.
Similarly, electron localization in the equal-arm loop chains in
the presence of the spin-orbit interaction can also be understood
in the same way. When an electron is moving in the upper arm of
the loop, the electron spinor will acquire a phase $kL$ and be
rotated by the spin-rotation operator $S_1$. When an electron is
moving in the lower arm of the loop, the electron spinor will also
acquire a phase $kL$ and be rotated by the spin-rotation operator
$S_2$. Here $S_1$ and $S_2$ depend on only the loop geometry and
the strength of the spin-orbit interaction. Total destruction of
electronic waves thus occurs when $(e^{ik L}S_1+e^{ikL}S_2)=0$ is
satisfied, or equivalently
$\sqrt{2[1+\cos(\Theta_{so})]}e^{ikL}S(k_{so})=0$ with
$\cos(\Theta_{so}) = \frac{1}{4}\texttt{Tr}(S_1 S_2^{-1}+ S_2
S_1^{-1})$. Therefore electron localization will occur in the
chain of equal-arm loops for $\cos\Theta_{so}=-1$.

The resemblance between the plots of the averaged transmission
probabilities can be explained as follows. As pointed in section
\ref{sec:scatter}, the transmission probability of a loop chain is
totally determined by the transfer matrix of a single loop
scatterer. The current conservation along with the single-channel
assumption for the quantum wires thus put strong restrictions on
the possible forms of the transfer matrix. Furthermore, both the
Aharonov-Bohm effect and the Aharonov-Casher effect play similar
roles in the electronic transmission problem. Electrons travelling
in the arms of the loops can acquire additional quantum phases due
to either the Aharonov-Bohm effect or the Aharonov-Casher effect.
Therefore, the transfer matrices for the loop scatterers discussed
in the paper also resemble each other. The resemblance between the
transfer matrices thus leads to the resemblance between the plots
of the averaged transmission probabilities.

\begin{acknowledgments}
This work was supported in part by National Science Council of
Taiwan.
\end{acknowledgments}
\bibliography{Ref}

\begin{thebibliography}{11}
\expandafter\ifx\csname natexlab\endcsname\relax\def\natexlab#1{#1}\fi
\expandafter\ifx\csname bibnamefont\endcsname\relax
  \def\bibnamefont#1{#1}\fi
\expandafter\ifx\csname bibfnamefont\endcsname\relax
  \def\bibfnamefont#1{#1}\fi
\expandafter\ifx\csname citenamefont\endcsname\relax
  \def\citenamefont#1{#1}\fi
\expandafter\ifx\csname url\endcsname\relax
  \def\url#1{\texttt{#1}}\fi
\expandafter\ifx\csname urlprefix\endcsname\relax\def\urlprefix{URL }\fi
\providecommand{\bibinfo}[2]{#2}
\providecommand{\eprint}[2][]{\url{#2}}

\bibitem[{\citenamefont{Vidal et~al.}(1998)\citenamefont{Vidal, Mosseri, and
  Dou\c{c}ot}}]{ABcage}
\bibinfo{author}{\bibfnamefont{J.}~\bibnamefont{Vidal}},
  \bibinfo{author}{\bibfnamefont{R.}~\bibnamefont{Mosseri}}, \bibnamefont{and}
  \bibinfo{author}{\bibfnamefont{B.}~\bibnamefont{Dou\c{c}ot}},
  \bibinfo{journal}{Phys. Rev. Lett.} \textbf{\bibinfo{volume}{81}},
  \bibinfo{pages}{5888} (\bibinfo{year}{1998}).

\bibitem[{\citenamefont{Aharonov and Bohm}(1959)}]{ABeffect}
\bibinfo{author}{\bibfnamefont{Y.}~\bibnamefont{Aharonov}} \bibnamefont{and}
  \bibinfo{author}{\bibfnamefont{D.}~\bibnamefont{Bohm}},
  \bibinfo{journal}{Phys. Rev.} \textbf{\bibinfo{volume}{115}},
  \bibinfo{pages}{485} (\bibinfo{year}{1959}).

\bibitem[{\citenamefont{\textit{et al}.}(1999)}]{ABcageSuper}
\bibinfo{author}{\bibfnamefont{C.~C.~A.} \bibnamefont{\textit{et al}.}},
  \bibinfo{journal}{Phys. Rev. Lett.} \textbf{\bibinfo{volume}{83}},
  \bibinfo{pages}{5102} (\bibinfo{year}{1999}).

\bibitem[{\citenamefont{Naud et~al.}(2001)\citenamefont{Naud, Faini, and
  Mailly}}]{ABcageMetal}
\bibinfo{author}{\bibfnamefont{C.}~\bibnamefont{Naud}},
  \bibinfo{author}{\bibfnamefont{G.}~\bibnamefont{Faini}}, \bibnamefont{and}
  \bibinfo{author}{\bibfnamefont{D.}~\bibnamefont{Mailly}},
  \bibinfo{journal}{Phys. Rev. Lett.} \textbf{\bibinfo{volume}{86}},
  \bibinfo{pages}{5104} (\bibinfo{year}{2001}).

\bibitem[{\citenamefont{Vidal et~al.}(2001)\citenamefont{Vidal, Butaud,
  Dou\c{c}ot, and Mosseri}}]{disorder}
\bibinfo{author}{\bibfnamefont{J.}~\bibnamefont{Vidal}},
  \bibinfo{author}{\bibfnamefont{P.}~\bibnamefont{Butaud}},
  \bibinfo{author}{\bibfnamefont{B.}~\bibnamefont{Dou\c{c}ot}},
  \bibnamefont{and} \bibinfo{author}{\bibfnamefont{R.}~\bibnamefont{Mosseri}},
  \bibinfo{journal}{Phys. Rev. B} \textbf{\bibinfo{volume}{64}},
  \bibinfo{pages}{155306} (\bibinfo{year}{2001}).

\bibitem[{\citenamefont{Vidal et~al.}(2000{\natexlab{a}})\citenamefont{Vidal,
  Dou\c{c}ot, Mosseri, and Butaud}}]{delocalization}
\bibinfo{author}{\bibfnamefont{J.}~\bibnamefont{Vidal}},
  \bibinfo{author}{\bibfnamefont{B.}~\bibnamefont{Dou\c{c}ot}},
  \bibinfo{author}{\bibfnamefont{R.}~\bibnamefont{Mosseri}}, \bibnamefont{and}
  \bibinfo{author}{\bibfnamefont{P.}~\bibnamefont{Butaud}},
  \bibinfo{journal}{Phys. Rev. Lett.} \textbf{\bibinfo{volume}{85}},
  \bibinfo{pages}{3906} (\bibinfo{year}{2000}{\natexlab{a}}).

\bibitem[{\citenamefont{Vidal et~al.}(2000{\natexlab{b}})\citenamefont{Vidal,
  Montambaux, and Dou\c{c}ot}}]{transport}
\bibinfo{author}{\bibfnamefont{J.}~\bibnamefont{Vidal}},
  \bibinfo{author}{\bibfnamefont{G.}~\bibnamefont{Montambaux}},
  \bibnamefont{and}
  \bibinfo{author}{\bibfnamefont{B.}~\bibnamefont{Dou\c{c}ot}},
  \bibinfo{journal}{Phys. Rev. B} \textbf{\bibinfo{volume}{62}},
  \bibinfo{pages}{R16294} (\bibinfo{year}{2000}{\natexlab{b}}).

\bibitem[{\citenamefont{Bercioux et~al.}(2004)\citenamefont{Bercioux,
  Governale, Cataudella, and Ramaglia}}]{RashbaNetwork}
\bibinfo{author}{\bibfnamefont{D.}~\bibnamefont{Bercioux}},
  \bibinfo{author}{\bibfnamefont{M.}~\bibnamefont{Governale}},
  \bibinfo{author}{\bibfnamefont{V.}~\bibnamefont{Cataudella}},
  \bibnamefont{and} \bibinfo{author}{\bibfnamefont{V.~M.}
  \bibnamefont{Ramaglia}}, \bibinfo{journal}{Phys. Rev. Lett.}
  \textbf{\bibinfo{volume}{93}}, \bibinfo{pages}{056802}
  (\bibinfo{year}{2004}).

\bibitem[{\citenamefont{Bercioux et~al.}(2005)\citenamefont{Bercioux,
  Governale, Cataudella, and Ramaglia}}]{RashbaNetDetail}
\bibinfo{author}{\bibfnamefont{D.}~\bibnamefont{Bercioux}},
  \bibinfo{author}{\bibfnamefont{M.}~\bibnamefont{Governale}},
  \bibinfo{author}{\bibfnamefont{V.}~\bibnamefont{Cataudella}},
  \bibnamefont{and} \bibinfo{author}{\bibfnamefont{V.~M.}
  \bibnamefont{Ramaglia}}, \bibinfo{journal}{Phys. Rev. B.}
  \textbf{\bibinfo{volume}{72}}, \bibinfo{pages}{075305}
  (\bibinfo{year}{2005}).

\bibitem[{\citenamefont{Aharonov and Casher}(1984)}]{ACeffect}
\bibinfo{author}{\bibfnamefont{Y.}~\bibnamefont{Aharonov}} \bibnamefont{and}
  \bibinfo{author}{\bibfnamefont{A.}~\bibnamefont{Casher}},
  \bibinfo{journal}{Phys. Rev. Lett.} \textbf{\bibinfo{volume}{53}},
  \bibinfo{pages}{319} (\bibinfo{year}{1984}).

\bibitem[{\citenamefont{Moln\'{a}r et~al.}(2005)\citenamefont{Moln\'{a}r,
  Vasilopoulos, and Peeters}}]{RashbaMagRings}
\bibinfo{author}{\bibfnamefont{B.}~\bibnamefont{Moln\'{a}r}},
  \bibinfo{author}{\bibfnamefont{P.}~\bibnamefont{Vasilopoulos}},
  \bibnamefont{and} \bibinfo{author}{\bibfnamefont{F.~M.}
  \bibnamefont{Peeters}}, \bibinfo{journal}{Phys. Rev. B.}
  \textbf{\bibinfo{volume}{72}}, \bibinfo{pages}{075330}
  (\bibinfo{year}{2005}).

\end{thebibliography}

\newpage

\end{document}